\documentclass[12pt,psfig]{article}
\usepackage{amssymb}
\usepackage{amsmath,amsthm}
\usepackage{amsfonts}
\usepackage{graphicx}
\usepackage{braket}
\usepackage{graphics}
\usepackage{indentfirst}
\numberwithin{equation}{section}

\usepackage{hyperref}

\begin{document}
\begin{center}\Large\textbf{Closed String
Radiation from the Interaction of
the Moving D$p$-branes with Background Fields}
\end{center}
\vspace{0.75cm}
\begin{center}
\large{Hamidreza  Daniali and \large Davoud Kamani}
\end{center}
\begin{center}
\textsl{\small{Department of Physics, Amirkabir University of
Technology (Tehran Polytechnic), Iran \\
P.O.Box: 15875-4413 \\
e-mails: hrdl@aut.ac.ir , kamani@aut.ac.ir \\}}
\end{center}
\vspace{0.5cm}

\begin{abstract}

At first, we derive a boundary state, associated
with a moving D$p$-brane,
in the presence of the Kalb-Ramond field and a $U(1)$
gauge potential. Then, from the
interaction of such D$p$-branes,
we obtain the radiation amplitude for a massless
closed string. The behavior of the radiation amplitude for
the large distance of the branes will be studied.
Our calculations is in the framework of the
bosonic string theory.

\end{abstract}


\newpage

\section{Introduction}

By adding various background and
internal fields to the D$p$-branes
the most general boundary states
corresponding to them,
as well as their interactions,
have been widely studied
\cite{1}-\cite{24}. Precisely, the D-branes
with the nonzero background fields
exhibit some appealing properties, e.g.,:
forming various bound states \cite{25,26},
generating fundamental strings
between the branes \cite{27},
loop corrections to the string
equations of motion \cite{28},
scattering of closed strings
from the branes \cite{29}. However,
since the D-branes are physical objects,
it also is necessary to comprise their dynamics
\cite{1, 2, 5, 30, 31}.
Thus, the D$p$-branes with the nonzero
background fields and dynamics (dressed-dynamical branes)
prominently reveal a variety of novel properties.

Closed string radiation can occur from the
D-branes with various configurations. For example,
closed string radiation from a single unstable
brane in the presence of the background fields
has been studied \cite{32}-\cite{36}.
Besides, the supersymmetric version of the
closed string radiation has been also
investigated \cite{37}.
Another interesting configuration
is the closed string radiation
from the interacting branes. This kind of radiation
has been studied only in some special setups
\cite{38}, \cite{39}, \cite{40}, \cite{41}.

The background field and the internal
gauge potential, accompanied by
the transverse motions of the branes,
motivated and stimulated us to
examine the simultaneous effects
of these effective factors on
the closed string radiation. Consequently,
we obtain a generalized radiation amplitude.
In other words, this form of the radiation
amplitude obviously comprises more adjustable parameters.
Thus, using the boundary state formalism,
we shall investigate the massless
closed string radiation, generated
by the interaction of two parallel D$p$-branes.
The branes have the transverse motions
and have been dressed by the Kalb-Ramond
field and the $U(1)$ gauge potentials.
Hence, we insert a suitable vertex operator into the
worldsheet of the closed string,
exchanged between the branes.
We restrict ourselves in the eikonal approximation
in which the recoil of the branes, due to the
string radiation, is ignored.
Since this paper is closely related to the
untwisted sector of Ref. \cite{42},
the features of some formulas
of this study naturally possess resemblances
with those of the mentioned reference.
Our calculations is in the
context of the bosonic string theory.

This paper is organized as follows. In Sec. \ref{200},
we shall introduce a boundary state corresponding to a
dressed-dynamical D$p$-brane.
In Sec. \ref{301}, we shall compute the amplitude of the
closed string radiation, produced by
the interaction of two dressed-dynamical
parallel D$p$-branes. In Sec. \ref{302},
the radiation amplitude
will be rewritten for the large distance branes.
Sec. \ref{400} is devoted to the conclusions.

\section{Dressed-dynamical D$p$-brane: the boundary state}
\label{200}

To establish the boundary state, associated with a
D$p$-brane with background fields,
the following sigma-model action for a
closed string is considered
\begin{equation}
\label{2.1}
S = -\dfrac{1}{4 \pi \alpha^\prime} \int_\Sigma
d^2 \sigma \left(\sqrt{-h} h^{AB} \eta_{\mu\nu}
+ \epsilon^{AB} B_{\mu\nu} \right)
\partial_A X^\mu \partial_B X^\nu +
\dfrac{1}{2\pi \alpha^\prime} \int_{\partial\Sigma}
d\sigma A_\alpha \partial_\sigma X^\alpha,
\end{equation}
where $\mu,\nu$ are the spacetime indices, and
$\alpha,\beta$ refer to the directions of the
brane worldvolume. We apply the
flat spacetime with the metric
$\eta_{\mu\nu}={\rm diag} (-1,1,\cdots, 1)$,
the flat worldsheet $h_{AB}=\eta_{AB}$
with $A,B\in \{0,1\}$,
a constant Kalb-Ramond field
$B_{\mu\nu}$, and the gauge potential in the
Landau gauge $A_\alpha = -\frac{1}{2}
F_{\alpha\beta} X^\beta$ with the constant field strength
$F_{\alpha\beta}$. The worldsheet of the closed
string is $\Sigma$, and $\partial \Sigma$ represents
its boundary.

By vanishing the variation of the action,
beside the equation of motion,
the following boundary state equations are extracted
\begin{eqnarray}
\left(\partial_\tau X^\alpha
+ \mathcal{F}^{\alpha}_{\ \ \beta}
\partial_\sigma
X^\beta \right)_{\tau=0} | B_x\rangle &=& 0 ,
\label{2.2}\\
\left(X^i - y^i\right)_{\tau=0} | B_x\rangle &=& 0,
\label{2.3}
\end{eqnarray}
where $\mathcal{F}_{\alpha\beta}
= F_{\alpha\beta} - B_{\alpha\beta}$
is the total field strength.
The set  $\{x^i| i = p+1,\cdots, 25\}$ indicates
the perpendicular directions to
the worldvolume of the D$p$-brane.
Thus, the parameters $y^i$ obviously
represent the location of the brane.

For imposing the transverse velocity
``$v$'' to the brane,
one should apply the boost transformations
on the boundary state equations.
Let the transverse direction
$x^{i_b}$ be the boost direction.
Hence, Eqs. \eqref{2.2} and \eqref{2.3} are
deformed as in the following
\begin{eqnarray}
&&\left[\partial_\tau (X^0 - v X^{i_b})
+ \mathcal{F}^0_{\ \ \bar{\alpha} }\partial_{\sigma}
X^{\bar{\alpha}}\right]_{\tau=0}
|B_x\rangle= 0, \nonumber \\
&&\left[\partial_\tau X^{\bar\alpha}
+ \gamma^2 \mathcal{F}^{\bar\alpha}_{\ \ 0}
\partial_\sigma (X^0 - v X^{i_b})
+ \mathcal{F}^{\bar\alpha}_{\ \ \bar\beta}\partial_\sigma
X^{\bar\beta}\right]_{\tau=0}
|B_x\rangle= 0,\nonumber \\
&&\left( X^{i_b} - v X^0 - y^{i_b}\right)_{\tau=0}
|B_x\rangle= 0,\nonumber \\
&& \left( X^{i} - y^{i}\right)_{\tau=0}
|B_x\rangle= 0 ,\qquad i \ne i_b
\label{2.4},
\end{eqnarray}
where we use $\{\bar{\alpha}\} = \{\alpha\} - \{0\}$ and
$\gamma = 1/\sqrt{1-v^2}$.

After applying the closed string mode expansion,
the boundary state equations \eqref{2.4} will be
rewritten in terms of the closed string
oscillators and the zero modes.
Therefore, by using the coherent state technique \cite{43},
the oscillating part of the boundary state
takes the form
\begin{eqnarray}
|B_x\rangle_{\rm osc} &=& \sqrt{- \det M}
\exp \left[ - \sum_{m=1}^{\infty} \left(\dfrac{1}{m}
\alpha^\mu_{-m} S_{\mu \nu}
\alpha^{\nu}_{-m}\right)\right]
{|0\rangle}_{\alpha }
\otimes{|0\rangle}_{\widetilde{\alpha }}\;.
\label{2.5}
\end{eqnarray}
The prefactor $\sqrt{-\det M}$
comes from the disk partition function \cite{28}.
The matrix $S_{\mu\nu}$ possesses the following definition
\begin{eqnarray}
S_{\mu\nu} &=& \Big(Q_{\lambda\lambda'}
=(M^{-1}N)_{\lambda\lambda'}
|_{\lambda,\lambda' \in \{\alpha,i_b\}}\;
,\;-\delta_{ij}|_{i,j \ne i_b}  \Big),
\label{2.6}
\end{eqnarray}
where the matrices $M$ and $N$ are given by
\begin{eqnarray}
M^0_{\ \lambda} &=& \gamma\left(\delta^0_{\ \lambda} -
v \delta^{i_b}_{\ \lambda} - \mathcal{F}^0_{\ \bar\alpha}
\delta^{\bar\alpha}_{\ \lambda}\right),
\nonumber\\
M^{\bar\alpha}_{\ \lambda} &=&
\delta^{\bar\alpha}_{\ \lambda}
-\gamma^2 \mathcal{F}^{\bar\alpha}_{\ 0}(\delta^{0}_{\ \lambda}
-v\delta^{i_b}_{\ \lambda})
-\mathcal{F}^{\bar\alpha}_{\;\;\;\bar\beta}
\delta^{\bar\beta}_{\ \lambda},
\nonumber\\
M^{i_b}_{\ \lambda} &=& \delta^{i_b}_{\ \lambda}
- v \delta^{i_b}_{\ \lambda},
\label{2.7}
\end{eqnarray}
\begin{eqnarray}
N^0_{\ \lambda} &=& \gamma\left(\delta^0_{\ \lambda}
- v \delta^{i_b}_{\ \lambda}
+ \mathcal{F}^0_{\ \bar\alpha}
\delta^{\bar\alpha}_{\ \lambda}\right),
\nonumber\\
N^{\bar\alpha}_{\ \lambda} &=&
\delta^{\bar\alpha}_{\ \lambda}
+ \gamma^2 \mathcal{F}^{\bar\alpha}_{\ 0}
(\delta^{0}_{\ \lambda}
- v\delta^{i_b}_{\ \lambda})
- \mathcal{F}^{\bar\alpha}_{\;\;\;\bar\beta}
\delta^{\bar\beta}_{\ \lambda},
\nonumber\\
N^{i_b}_{\ \lambda} &=& -\delta^{i_b}_{\ \lambda}
+ v \delta^{i_b}_{\ \lambda}.
\label{2.8}
\end{eqnarray}
Using the well-known commutation relations
in the quantum mechanics and also the
integral version of the Dirac delta function,
we acquire the zero-mode part of the boundary state
in the form
\begin{equation}
\label{2.9}
|B_x\rangle_0= \int_{-\infty}^{+\infty}
\prod_{i = p+1}^{25} \dfrac{{\rm d} p^i}{2\pi}
\exp\left(-ip^i y_i\right)
\prod^{25}_{\mu=0}|p^\mu \rangle,
\end{equation}
in which $p^0 =vp^{i_b}$, $p^{\bar \alpha} = 0$,
and all $p^i$s (including $p^{i_b}$) are nonzero.

The following direct product clarifies the total
boundary state, corresponding to the D$p$-brane, in the
context of the bosonic string theory
\begin{equation}
\label{2.10}
|B_x\rangle_{\rm tot}= \frac{T_p}{2}
|B_x\rangle_{\rm osc} \otimes
|B_x\rangle_{0} \otimes |B_{\rm g}\rangle,
\end{equation}
where $|B_{\rm g}\rangle$ is the
known boundary state of the
conformal ghosts, and $T_p$ is the brane tension.

\section{Closed string radiation}
\label{300}

\subsection{The branes with arbitrary distance}
\label{301}

The interaction of two D$p$-branes in the closed
string channel
occurs by exchanging a closed string between two
boundary states, corresponding to the branes.
Let $\sigma$ denote the periodic
coordinate of the string worldsheet
$ 0\le \sigma\le \pi$,
and $\tau$ as the coordinate along the length of
it $ 0\le \tau\le t$.
The emission of a closed string state is described via
an appropriate vertex operator $V(\tau, \sigma)$.
Therefore, we require to calculate
the following amplitude \cite{38},
\begin{equation}
\label{3.1}
\mathcal{A}
= \int_{0}^{\infty} {\rm d}t \int_{0}^{t} {\rm d}\tau \
_{\rm tot}\langle B_1|
e^{-tH} V(\tau, \sigma)
|B_2\rangle_{\rm tot},
\end{equation}
where $H$ is the total closed string Hamiltonian
(including the matter and ghost parts)
\begin{eqnarray}
H &=& H_{\rm g} + \alpha^\prime p^2
+ 2 \sum_{n=1}^\infty \left( \alpha_{-n}.\alpha_{n}
+ \tilde{\alpha}_{-n}.\tilde{\alpha}_{n}\right)- 4.
\label{3.2}
\end{eqnarray}
Since the fields and velocities
of the interacting branes can be different, we employ
the subscripts (1) and (2) for the first and
second branes, respectively.

The vertex operator in Eq. \eqref{3.1} for
the massless closed string states
has the form $V(z, \bar{z}) = \epsilon_{\mu\nu}
\partial X^\mu \bar{\partial}X^\nu e^{ip\cdot X}$,
in which $\epsilon_{\mu\nu}$ is
the polarization tensor, and
$z=\sigma +i \tau$ and $\partial = \partial_z$.
We shall use $k_1$ and $k_2$
as the momenta of the emitted closed string from
the first brane and the absorbed one by
the second brane, respectively. Besides, ``$p$''
indicates the momentum of the radiated closed string,
in which $k_1 = k_2 + p$.
With these notations, and after a heavy calculation,
the radiation amplitude takes the feature
\begin{eqnarray}
\label{3.3}
\mathcal{A}&=& \epsilon_{\mu\nu}
\dfrac{T_p^2 }{4 (2\pi)^{24-2p}}
\frac{\prod^p_{\bar \alpha =1}\delta (p^{\bar \alpha})}
{|v_1- v_2|} \int_0^{\infty}
{\rm d}t' \int_0^\infty {\rm d}\tau
\int_{-\infty}^{+\infty}
\prod_{i \ne i_b}^{25} {\rm d}k_1^i e^{i k_1^i b_i}
\nonumber\\
&\times& e^{- t' \alpha'k_1^2 }
e^{- \tau \alpha'k_2^2  }
e^{4(t'+\tau)} \mathcal{Z}_{\rm osc,g}
\langle e^{ip\cdot X_{\rm osc}}
\rangle \Big[ \langle\partial X^\mu \bar{\partial}
X^\nu \rangle_{\rm osc}
\nonumber\\
&-& p_\gamma p_\eta \langle \partial X^\mu
X^\gamma \rangle_{\rm osc}\langle \bar\partial X^\nu
X^\eta \rangle_{\rm osc}
- i \alpha' k_1^\nu p_\gamma \langle \bar\partial X^\mu
X^\gamma \rangle_{\rm osc}
\nonumber\\
&+&i\alpha' k_1^\mu p_\gamma \langle \partial X^\nu
X^\gamma \rangle_{\rm osc}
- \alpha'^2 k_1^\mu k_1^\nu \Big].
\end{eqnarray}
The partition function $\mathcal{Z}_{\rm osc,g}$
and the general form of the correlators are given by
\begin{eqnarray}
\label{3.4}
\mathcal{Z}_{\rm osc,g} &=& \ _{\rm osc}\langle B_1|
e^{-t H(X_{\rm osc})}|B_2\rangle_{\rm osc}\;
\langle B_{\rm g}|e^{-t H_{\rm g} }
|B_{\rm g}\rangle,
\nonumber\\
\langle \mathcal{O}(\sigma , \tau)\rangle_{\rm osc}
&=& \dfrac{_{\rm osc}\langle B_1|
e^{-t H_{\rm osc} }\mathcal{O} (\sigma , \tau)
|B_2\rangle_{\rm osc}}{_{\rm osc}
\langle B_1| e^{-t H_{\rm osc} }
|B_2\rangle_{\rm osc}}\;.
\end{eqnarray}
The indices $\gamma,\eta$ show the spacetime directions.
In addition, we defined the impact
parameter $b^i = y^i_1 - y^i_2$.
Note that, by using the Wick's rotation
$\tau \to -i\tau$,
we introduced another proper time $t' = t - \tau$. The
presence of the delta function clearly elaborates that the
radiated closed string moves perpendicular to the brane.

Applying the Hamiltonian \eqref{3.2} gives
the partition function as in the following
\begin{eqnarray}
\mathcal{Z}_{\rm osc,g} &=&
\sqrt{\det(M_1 M_2)} \prod_{n=1}^{\infty}
\bigg{\{}\left[{\rm det} \left(\mathbf{1}
- {\mathcal{S}} \ e^{-4nt}\right)\right]^{-1}
\left({\mathbf 1}- e^{-4nt} \right)^{2}\bigg{\}},
\label{3.5}
\end{eqnarray}
where ${\mathcal{S}} \equiv S_1^{\rm T} S_2$.
For the single derivative correlators in
Eq. \eqref{3.3} we receive
\begin{eqnarray}
\langle \partial X^\mu X^\nu \rangle_{\rm osc}
&=& i \alpha^\prime \Bigg\{\eta^{\mu\nu}
\sum_{n=1}^{\infty} \text{Tr} \left( \dfrac{{\mathbf 1}
+ {\mathcal{S}} e^{-4nt} }{{\mathbf 1} - {\mathcal{S}}
e^{-4nt}}\right)  - \sum_{n=1}^{\infty}
\left(S_2 S_1 \right)^{\nu\mu}
\text{Tr} \left( \dfrac{ {\mathcal{S}}
e^{-4nt} }{{\mathbf 1} -
{\mathcal{S}} e^{-4nt}}\right)
\nonumber\\
&-& \sum_{n=0}^{\infty} S^{ \mu\nu}_{2}
\text{Tr} \Bigg[ \dfrac{{\mathcal{S}} e^{-4(nt+t')}}
{{\mathbf 1}- {\mathcal{S}} e^{-4(nt+t')}}
\left( {\mathbf 1} + \dfrac{1}{{\mathbf 1}- {\mathcal{S}}
e^{-4(nt+t')}}\right)\Bigg]
\nonumber\\
&+& \sum_{n=0}^{\infty}
\left(S_{1}\right)^{\nu\mu}
\text{Tr} \Bigg[ \dfrac{{\mathcal{S}} e^{-4(nt+\tau)}}
{{\mathbf 1}- {\mathcal{S}} e^{-4(nt+\tau)}}
\left({\mathbf 1} + \dfrac{1}
{{\mathbf 1}- {\mathcal{S}} e^{-4(nt+\tau)}}
\right)\Bigg]\Bigg\} .
\label{3.6}
\end{eqnarray}
One can obviously deduce that $\langle \partial X^\mu  X^\nu
\rangle_{\rm osc} = - \langle \bar{\partial} X^\mu X^\nu
\rangle_{\rm osc}$.
Besides, the two-derivative correlators can be computed by
taking the derivative of Eq. \eqref{3.6}.

The exponential correlator in Eq. \eqref{3.3}
can be computed via the Cumulant
expansion in the eikonal approximation. In this
method the recoil of the branes, due to
the closed string emission, is
neglected and the branes are assumed to
move in straight trajectories.
Let us assume that the radiated string
has a small momentum.
Therefore, in the Cumulant expansion,
we should compute the factor
$\exp\left(-\frac{1}{2} p_\mu p_\nu \langle
X^\mu X^\nu\rangle_{\rm osc}\right)$.
Finally, we obtain
$\langle e^{ip\cdot X_{\rm osc}}\rangle$
in terms of the following determinants
\begin{eqnarray}
\langle e^{ip\cdot X_{\rm osc}} \rangle
&\approx & \prod_{n=1}^{\infty}
\det \left({\mathbf 1}- {\mathcal{S}}
e^{-4nt}\right)^{\frac{\alpha^\prime}{2n}
p_\mu p_\nu \left(S_2S_1\right)^{\mu\nu}}
\nonumber\\
&\times& \prod_{n=0}^{\infty}
\det\left[ \left({\mathbf 1}- {\mathcal{S}}e^{-4(nt+t')}
\right)^{-\frac{\alpha^\prime}{2(n+1)}p_\mu p_\nu
S^{ \mu\nu}_{2}}\left({\mathbf 1}
- {\mathcal{S}} e^{-4(nt+\tau)}
\right)^{- \frac{\alpha^\prime}{2(n+1)}
p_\mu p_\nu S_1^{\mu\nu}}\right]
\nonumber\\
&\times& \prod_{n=0}^{\infty} \det \left\{ \exp
\left[ \frac{\alpha^\prime p_\mu p_\nu
S^{ \mu\nu}_{2}}{2(n+1)}  \left({\mathbf 1}- {\mathcal{S}}
e^{-4(nt+t')}\right)^{-1} \right] \right\}
\ \ \
\nonumber\\
&\times& \prod_{n=0}^{\infty} \det \left\{ \exp \left[
\frac{\alpha^\prime p_\mu p_\nu
S_1^{\mu\nu}}{2(n+1)}
\left({\mathbf 1}- {\mathcal{S}}
e^{-4(nt+\tau)}\right)^{-1} \right]
\right\} .
\label{3.7}
\end{eqnarray}
Putting all correlators into Eq. \eqref{3.3},
we obtain the radiation amplitude
of the radiated closed string.
Since the amplitude is very long, we do not
write the explicit form of it.

\subsection{The branes with a large distance}
\label{302}

Here we consider the closed string radiation
from the branes with a large distance.
When the branes separation is large,
they effectively interact
via the massless states of the closed string spectrum.
A sufficiently large time is equivalent to a large
inter-brane separation.
However, the limit $t\rightarrow\infty$ should be merely
applied to the oscillating portion
of the amplitude, i.e., the partition
functions and correlators.

For simplification, let us impose the condition
${\mathcal{S}}=\mathbf{1}$ to our formulation,
which restricts the fields and velocities of the branes.
Thus, the large distance limit (LDL),
accompanied by the condition ${\mathcal{S}}=\mathbf{1}$,
specify the partition function as
$\mathcal{Z}_{\rm osc,g}
|_{\rm LDL}^{{\mathcal{S}} =\mathbf{1}}
= \sqrt{\det(M_1 M_2)}$,
and the following features for the correlators
\begin{eqnarray}
\langle \partial X^\mu X^\nu\rangle_{\rm osc}
|_{\rm LDL}^{{\mathcal{S}}
= \mathbf{1}}  &=& 26 i \alpha'
\left\{ -\frac{1}{2}\eta^{\mu\nu}
- S_2^{\mu\nu} \left[\frac{e^{-4t'}}{1 - e^{-4t'}}
\left( 1+ \frac{1}{1- e^{-4t'}}\right)\right]\right.
\nonumber\\
&+&\left. \left(S_1\right)^{\nu\mu}
\left[ \frac{e^{-4\tau}}{1- e^{-4\tau}}
\left( 1 + \frac{1}{1- e^{-4\tau}}\right)\right]\right\},
\label{3.8}
\end{eqnarray}
\begin{eqnarray}
\langle e^{ip\cdot X_{\rm osc}} \rangle
|_{\rm LDL}^{{\mathcal{S}}= \mathbf{1}}
&= &  \left[\Big(1 - e^{-4\tau}\Big)^{p_\mu p_\nu
S_1^{\mu\nu}}\Big(1 -  e^{-4t'}\Big)^{p_\mu p_\nu
S_2^{\mu\nu}}\right]^{-13 \alpha'}
\nonumber\\
&\times &\exp\bigg{\{}13\alpha'
p_\mu p_\nu\bigg{(}\frac{
S_1^{\mu\nu}}{1- e^{-4\tau}} +\frac{S_2^{ \mu\nu}}
{1- e^{-4t'}}
\nonumber\\
&+&\left(S_1 +S_2\right)^{\mu\nu}
\left(\gamma_{\rm EM}-1\right)\bigg{)}\bigg{\}}.
\label{3.9}
\end{eqnarray}
The Euler-Mascheroni number
$\gamma_{\rm EM}= 0.577 \cdots $ was entered
via a regularization scheme.
According to Eq. \eqref{3.6}, one may conveniently
recast the two-derivative correlator in
terms of the one-derivative correlators. Using
$\bar\partial = \frac{i}{2} (\partial_\tau|_{t'}
- \partial_{t'}|_{\tau} )$, and
applying the integration by part, one finds
\begin{eqnarray}
\langle \partial X^\mu \bar{\partial} X^\nu
\rangle_{\rm osc}|_{\rm LDL}^{{\mathcal{S}}
= \mathbf{1}}
= - \langle \partial X^\mu
X^\nu\rangle_{\rm osc}|_{\rm LDL}^{{\mathcal{S}}
= \mathbf{1}}
\left[ \langle p\cdot\partial X
p\cdot X\rangle_{\rm osc}|_{\rm LDL}^{{\mathcal{S}}
= \mathbf{1}}  + \dfrac{i \alpha^\prime}{2}
(k^2_1 - k^{ 2}_2)\right].
\label{3.10}
\end{eqnarray}

Replacing Eqs. \eqref{3.8}, \eqref{3.9}
and \eqref{3.10} into Eq. \eqref{3.3} gives
the radiation amplitude of the closed string
via the large distance branes.
In fact, the form of this amplitude also is very long.
Hence, we do not explicitly write it.
The explicit form of the amplitude for the radiation of
graviton, Kalb-Ramond and dilaton states
has been given by Eq. \eqref{3.15}.

We should mention that through
the calculations a non-integral term appeared.
It is required to avoid the non-integral
term to possess physical features.
Therefore, in the exponential correlator,
some undesirable surface terms at $\tau ,\; t' = 0$
have been excluded by analytic continuation.
This implies that the quantities
$p_\mu p_\nu S_1^{\mu\nu}$ and
$p_\mu p_\nu S_2^{ \mu\nu}$ should be negative.

\subsubsection{The radiation amplitude
of the graviton, Kalb-Ramond and dilaton states}

In Eq. \eqref{3.3}
we should determine two integrals on the proper times.
Using the integration by part,  
the following equivalence relations are
obtained \cite{38}, \cite{39},
\begin{eqnarray}
\ \frac{e^{-4\tau}}{1
- e^{-4\tau}} \left( 1 + \frac{1}{1
- e^{-4\tau}}\right) &\doteq&
\frac{4-\alpha'k_2^2}{52\alpha' p_\mu p_\nu S_1^{\mu \nu}},
\nonumber\\
\frac{e^{-4t'}}{1 - e^{-4t'}} \left( 1
+ \frac{1}{1- e^{-4t'}}\right)
&\doteq& \frac{4-\alpha'k_1^2}
{52\alpha' p_\mu p_\nu S_2^{ \mu \nu}}.
\label{3.11}
\end{eqnarray}
Now let us apply the low energy limit
assumption, in which the momentum of
the radiated string is small. Thus, the solution
of the proper times integrals is given by
\begin{equation}
\int_0^{\infty} {\rm d}t' \int_0^\infty
{\rm d}\tau \ e^{- \alpha'k_1^2 t'}
e^{- \alpha'k_2^2 \tau}e^{4(t'+\tau)}
\langle e^{ip\cdot X_{\rm osc}}
\rangle|_{\rm LDL}^{{\mathcal{S}}
= \mathbf{1}} \approx \Big[(\alpha' k_1^2 -4)
(\alpha' k_2^2 -4)\Big]^{-1}.
\label{3.12}
\end{equation}

For separating the string radiation to the graviton,
Kalb-Ramond and dilaton, one should
apply their corresponding polarization tensors
\begin{eqnarray}
{\rm Graviton:}&&\epsilon^{\rm g}_{\mu\nu} =
\epsilon^{\rm g}_{\nu\mu}, \quad
\epsilon^{\rm{g} \ \mu}_\mu = 0,
\nonumber\\
{\rm Kalb-Ramond:}&&\epsilon^{\rm KR}_{\mu\nu}
= -\epsilon^{\rm KR}_{\nu\mu},
\nonumber\\
{\rm Dilaton:}&& \epsilon^{\rm D}_{\mu\nu}
= \frac{1}{\sqrt{24}}
\left( \eta_{\mu\nu} - p_\mu \bar p_\nu
- p_\nu \bar p_\mu\right),
\;\; \bar p^2  =0, \;\; p \cdot\bar p =1.
\label{3.13}
\end{eqnarray}
The polarization tensors should satisfy
the condition $p^\mu \epsilon_{\mu\nu} = 0$.
To maintain the generality of the formulation,
we do not impose extra conditions to
the polarization tensors.

For writing the radiation amplitude in a desirable
form, let us apply the constant shifts on the momenta, i.e.
$k^\alpha_{1,2}-l^\alpha \rightarrow k^\alpha_{1,2}$,
in which the internal vector
$ l^\alpha$ satisfies the following conditions
\begin{eqnarray}
k_1 \cdot l = k_2\cdot l =0\;,\;\;\;\;
l^2= -\frac{4}{\alpha^\prime}.
\end{eqnarray}

The general structure of the three amplitudes
for the graviton, Kalb-Ramond and dilaton
radiation possesses the following feature
\begin{eqnarray}
\mathcal{A}|_{\rm LDL}^{{\mathcal{S}}
= \mathbf{1}} = &=& \dfrac{T_p^2
\sqrt{\det(M_1 M_2)}}{4(2\pi)^{24-2p}}
\frac{1}{|v_1^{i_b}- v_2^{i_b}|}
\prod_{\bar\alpha =1}^p \delta
(p^{\bar\alpha})
\nonumber\\
&\times& \int_{-\infty}^{+\infty}
\prod_{i \ne i_b}^{25} dk_1^i e^{i k_1^i b_i}
\left\{\frac{\mathbf{F}}{k_1^2 }
+ \frac{\mathbf{M}}
{k_1^2  k^2_2} + \frac{\mathbf{S}}{k_2^2 }
\right\}.
\label{3.15}
\end{eqnarray}
The explicit forms of the set
$\{\mathbf{F}, \mathbf{M}, \mathbf{S} \}$
for each of the massless states are given by
\begin{eqnarray}
\mathbf{F}^\text{g} &=& \frac{1}{2 p_\mu p_\nu
S_1^{ \mu\nu}}
\Bigg\{\epsilon^\text{g}_{\alpha\beta}
\left[ 26 \left(p^\alpha
+ \frac{1}{13}k_1^\alpha\right)\left(p^\beta
+ p_\theta Q_1^{T \beta\theta}\right)
- \frac{1}{2} Q_1^{T \alpha\beta} (k_1^2
- k_2^2)\right]
\nonumber\\
&-& \frac{1}{2} \epsilon_i^{\text{g} i}
\left[ 50 p^i p_i
+ 24 p_\xi p_\theta Q_1^{\xi\theta}
+ 26 p^\alpha p_\alpha + (k_1^2
- k_2^2) \right]
\nonumber\\
&+& \epsilon_{ij}^\text{g} (2 k_1^i p^j
- 26 p^i p^j) - \frac{
k^{2}_2 }{2 p_\mu p_\nu  S_1^{ \mu\nu}}
\Bigg( \epsilon^{\rm g}_{\alpha\beta}
\bigg[ p_\xi p_\theta (Q_1^{T \alpha\beta}
Q_1^{ \xi\theta} - Q_1^{T \alpha\xi}
Q_1^{T \beta\theta})
\nonumber \\
&+& 2 p_\xi p^\beta Q_1^{T \alpha\xi}
+ p^ip_i Q_1^{T \alpha\beta}\bigg]
+ \epsilon^{\rm{g}}_{ij} \left[ \delta^{ij}
(p_\xi p_\theta Q_1^{ \xi\theta}
+ p^kp_k) - p^i p^j\right]
\Bigg)\Bigg\} ,
\end{eqnarray}
\begin{eqnarray}
\mathbf{F}^\text{KR} &=& \frac{1}{2 p_\mu p_\nu
S_1^{ \mu\nu}}\Bigg\{13 \epsilon^{\rm A}_{\alpha\beta}
p_\xi \left[ \left(p + \frac{1}{13}
k_1\right)^\alpha Q_1^{T \beta\xi}
+ \left(p + \frac{1}{13} k_1\right)^\beta
Q_1^{T \alpha\xi} \right]
\nonumber\\
&-& \frac{k^{2}_2 }{2 p_\mu p_\nu  S_1^{ \mu\nu}}
\epsilon^{\rm A}_{\alpha\beta}
\left[ p_\xi p_\theta (Q_1^{T \alpha\beta}
Q_1^{ \xi\theta} - Q_1^{T \alpha\xi} Q_1^{T \beta\theta})
+ p^ip_i Q_1^{T \alpha\beta}\right]\Bigg\},
\end{eqnarray}
\begin{eqnarray}
\mathbf{F}^{\rm D} &=& \frac{1}{2 \sqrt{24}
p_\mu p_\nu S_1^{ \mu\nu}}
\Bigg\{ \left\{ 26\left( p_\xi Q_{1 \  \alpha}^{T \xi}
+ p_\alpha\right)
\left( p + \frac{1}{13} k_1\right)^\alpha - \frac{1}{2}
Q_{1 \ \alpha}^{T \alpha} (k_1^2
- k_2^2 ) \right.
\nonumber \\
&-& (p_\alpha \bar p_\beta - p_\beta \bar p_\alpha)
\left [ 26\left( p_\xi Q_1^{T \beta\xi}
+ p^\beta\right)\left(p
+ \frac{1}{13}  k_1\right)^\alpha - \frac{1}{2}
Q_1^{T \alpha\beta} (k_1^2
- k_2^2 )\right]
\nonumber \\
&+& 2 k_1^i p_i - 26 p^i p_i - 2 p_i \bar p_j
\left(2 k_1^ip^j - 26 p^i p^j\right) - (p+1) \left(p_\xi
p_\theta Q_1^{ \xi \theta}
+ p^i p_i\right)
\nonumber \\
&+& 2 p_\alpha \bar p^\alpha
\left(p_\xi p_\theta
Q_1^{ \xi\theta} + p^i p_i\right)
+ (25- p - 2 p_i \bar p^i)
\left[ 26 p^i p_i +\frac{1}{2} (k_1^2
- k_2^2)\right.
\nonumber \\
&+& \left.\left. 13 p_\xi p_\theta
\left(Q_1^{ \xi\theta}
+ \eta^{\xi\theta}\right)
\right]\right\}
- \frac{k^{2}_2 }{2 p_\mu p_\nu  S_1^{ \mu\nu}}
\bigg\{ p_\xi p_\theta \left(Q_{1 \ \alpha}^{T \alpha }
Q_1^{ \xi \theta}
- Q_1^{T \alpha \xi } Q_{1 \alpha}^{T \ \theta}\right)
\nonumber \\
&+& 2 p_\xi p_\alpha Q_1^{ \alpha\xi}
+ p^i p_i Q_{1\ \alpha}^{T \alpha}
- (p_\alpha \bar p_\beta - p_\beta \bar p_\alpha)
\bigg[ p_\xi p_\theta \left(Q_1^{T \alpha\beta }
Q_1^{T \xi \theta}
- Q_1^{T \alpha \xi } Q_{1 }^{T \beta\theta}
\right)
\nonumber \\
&+& 2 p_\xi p^\beta Q_1^{T \alpha\xi}
+ p^i p_i Q_{1}^{T \alpha\beta}\bigg]
+ \left[(25-p)\left( p_\xi p_\theta
Q_1^{\xi \theta}
+ p^i p_i\right) - p^i p_i\right]
\nonumber \\
&-& 2 p_i \bar p^i p_\xi
p_\theta Q_1^{\xi\theta}\bigg\}\Bigg\} ,
\end{eqnarray}
\begin{eqnarray}
\mathbf{M}^\text{g} &=&
\epsilon_{\alpha\beta}^\text{g}
\left[ \frac{13}{2} \eta^{\alpha\beta} (k_1^2
- k_2^2 ) + (26 p - k_1)^\alpha k_1^\beta
- 169 p^\alpha p^\beta\right]
\nonumber \\
&+& 13 \ \epsilon_{ij}^\text{g} \left[ \left( k_1
- 13 p\right)^i \left( p - \frac{1}{13}k_1\right)^j
+ \frac{1}{2} \delta^{ij}  \right]
\nonumber \\
&+& 26 \ \epsilon^\text{g}_{\alpha i} \left( k_1
+ 13 p\right)^i \left( p
- \frac{1}{13}k_1\right)^\alpha ,
\end{eqnarray}
\begin{eqnarray}
\mathbf{M}^\text{KR} &=& 0 ,
\end{eqnarray}
\begin{eqnarray}
\mathbf{M}^{\rm D} &=& \frac{1}{\sqrt{24}}\left\{
\frac{13(p+1)}{2} (k_1^2 - k_2^2 )
+ \left(26 p +k_1\right)^\alpha k_{1 \alpha}
+ 169 p^\alpha p_\alpha \right.
\nonumber \\
&-&  2 p_\alpha \bar p_\beta \left( \frac{13}{2}
\eta^{\alpha\beta} (k_1^2  - k_2^2 )
+ (26 p + k_1)^\alpha k_1^\beta + 169
p^\alpha p^\beta \right)
\nonumber\\
&+& 13 \left[ \frac{25-p}{2}
(k_1^2  - k_2^2 )
 +  \left(p - \frac{1}{13} k_1\right)_i
 \left(k_1 - 13 p\right)^i \right]
\nonumber\\
 &+& [26 p_\alpha - 52 p_\alpha \bar p_\beta p^\beta
 +k_1^i (p_\alpha \bar p_i + p_i \bar p_\alpha)]
 \left( k_1 + 13 p\right)^\alpha \bigg\} ,
\end{eqnarray}
\begin{eqnarray}
\mathbf{S}^\text{g} &=& \frac{1}{2 p_\mu p_\nu
S_2^{ \mu\nu}}
\Bigg\{\epsilon^\text{g}_{\alpha\beta}
\left[ 26 \left(p^\alpha
+ \frac{1}{13}k_1^\alpha\right)\left(p^\beta
+ p_\theta Q_2^{ \beta\theta}\right)
- \frac{1}{2} Q_2^{ \alpha\beta} (k_1^2
- k_2^2)\right]
\nonumber \\
&-& \frac{1}{2} \epsilon_i^{\text{g} i}
\left[ 50 p^i p_i
+ 24 p_\xi p_\theta Q_2^{\xi\theta}
+ 26 p^\alpha p_\alpha + (k_1^2
- k_2^2) \right]
\nonumber \\
&+& \epsilon_{ij}^\text{g} (2 k_1^i p^j
- 26 p^i p^j) - \frac{
k^{2}_1 }{2 p_\mu p_\nu  S_2^{ \mu\nu}}
\Bigg( \epsilon^{\rm g}_{\alpha\beta}
\bigg[ p_\xi p_\theta (Q_2^{ \alpha\beta}
Q_2^{ \xi\theta} - Q_2^{ \alpha\xi}
Q_2^{ \beta\theta})
\nonumber \\
&+& 2 p_\xi p^\beta Q_2^{ \alpha\xi}
+ p^ip_i Q_2^{ \alpha\beta}\bigg]
+ \epsilon^{\rm{g}}_{ij} \left[ \delta^{ij}
(p_\xi p_\theta Q_2^{ \xi\theta}
+ p^kp_k) - p^i p^j\right]
\Bigg)\Bigg\} ,
\end{eqnarray}
\begin{eqnarray}
\mathbf{S}^\text{KR} &=& \frac{1}{2 p_\mu p_\nu
S_2^{ \mu\nu}}
\Bigg\{13 \epsilon^{\rm A}_{\alpha\beta}
p_\xi \left[ \left(p + \frac{1}{13}
k_1\right)^\alpha Q_2^{ \beta\xi}
+ \left(p + \frac{1}{13} k_1\right)^\beta
Q_2^{ \alpha\xi} \right]
\nonumber \\
&-& \frac{k^{2}_1 }{2 p_\mu p_\nu  S_2^{ \mu\nu}}
\epsilon^{\rm A}_{\alpha\beta}
\left[ p_\xi p_\theta (Q_2^{ \alpha\beta}
Q_2^{ \xi\theta} - Q_2^{ \alpha\xi} Q_2^{ \beta\theta})
+ p^ip_i Q_2^{ \alpha\beta}\right]\Bigg\} ,
\end{eqnarray}
\begin{eqnarray}
\mathbf{S}^{\rm D} &=& \frac{1}{2 \sqrt{24}
p_\mu p_\nu S_2^{ \mu\nu}}
\Bigg\{ \left\{ 26\left( p_\xi Q_{2 \  \alpha}^{ \xi}
+ p_\alpha\right)
\left( p + \frac{1}{13} k_1\right)^\alpha - \frac{1}{2}
Q_{2 \ \alpha}^{ \alpha} (k_1^2
- k_2^2 ) \right.
\nonumber \\
&-& (p_\alpha \bar p_\beta - p_\beta \bar p_\alpha)
\left [ 26\left( p_\xi Q_2^{ \beta\xi}
+ p^\beta\right)\left(p
+ \frac{1}{13}  k_1\right)^\alpha - \frac{1}{2}
Q_2^{ \alpha\beta} (k_1^2
- k_2^2 )\right]
\nonumber \\
&+& 2 k_1^i p_i - 26 p^i p_i - 2 p_i \bar p_j
\left(2 k_1^ip^j - 26 p^i p^j\right) - (p+1) \left(p_\xi
p_\theta Q_2^{ \xi \theta}
+ p^i p_i\right)
\nonumber \\
&+& 2 p_\alpha \bar p^\alpha
\left(p_\xi p_\theta
Q_2^{ \xi\theta} + p^i p_i\right)
+ (25- p - 2 p_i \bar p^i)
\left[ 26 p^i p_i +\frac{1}{2} (k_1^2
- k_2^2)\right.
\nonumber \\
&+& \left.\left. 13 p_\xi p_\theta
\left(Q_2^{ \xi\theta}
+ \eta^{\xi\theta}\right)
\right]\right\}
- \frac{k^{2}_1 }{2 p_\mu p_\nu  S_2^{ \mu\nu}}
\bigg\{ p_\xi p_\theta \left(Q_{2 \ \alpha}^{ \alpha }
Q_2^{ \xi \theta}
- Q_2^{ \alpha \xi } Q_{2 \alpha}^{ \ \theta}\right)
\nonumber \\
&+& 2 p_\xi p_\alpha Q_2^{ \alpha\xi}
+ p^i p_i Q_{2\ \alpha}^{ \alpha}
- (p_\alpha \bar p_\beta - p_\beta \bar p_\alpha)
\bigg[ p_\xi p_\theta \left(Q_2^{ \alpha\beta }
Q_2^{ \xi \theta}
- Q_2^{ \alpha \xi } Q_{2 }^{ \beta\theta}
\right)
\nonumber\\
&+& 2 p_\xi p^\beta Q_2^{ \alpha\xi}
+ p^i p_i Q_{2}^{ \alpha\beta}\bigg]
+ \left[(25-p)\left( p_\xi p_\theta
Q_2^{\xi \theta}
+ p^i p_i\right) - p^i p_i\right]
\nonumber\\
&-& 2 p_i \bar p^i p_\xi
p_\theta Q_2^{\xi\theta}\bigg\}\Bigg\} .
\end{eqnarray}

In the resultant amplitude \eqref{3.15},
the term with $\mathbf{M}$ refers
to the double-pole mechanism,
while each of the terms with $\mathbf F$
and $\mathbf S$ is the residue of a single-pole process.
We observe that the three $\mathbf{M}$s
completely are independent of the velocities and
fields of both branes, while the three $\mathbf{F}$s
(the three $\mathbf{S}$s) prominently depend on the
velocity and field of the first (the second) brane.
Therefore, for each of the massless strings,
the $\mathbf{M}$-term elucidates a
string radiation between the branes, but far
from both branes. The $\mathbf{F}$-term
($\mathbf{S}$-term) clarifies that
a massless string is emitted
by the second (the first)
brane, afterward it is absorbed
by the first (the second)
brane, then after moving as an
excited state, it re-decays by emitting
the final closed string on the first (second)
brane.

The non-vanishing values of $\mathbf{F}$,
$\mathbf{M}$ and $\mathbf{S}$
for the graviton and dilaton reveal that
these states can be radiated
in the three physical processes:
two emissions from each
of the interacting branes and
one from the middle points between them.
For the Kalb-Ramond state
we have $\mathbf{M}^{\rm KR}=0$.
This elaborates that the Kalb-Ramond production in
the middle points between branes is forbidden.
We should note that we introduced only one
vertex operator. Thus, we have only one radiated
closed string.

\section{Conclusions}
\label{400}

We introduced a boundary state, associated with
a moving D$p$-brane. The brane was
equipped with the Kalb-Ramond
background field and a $U(1)$
internal gauge potential in a specific gauge.
We extracted the radiation amplitude
of a massless closed string, which was created by
the interaction of two
parallel dressed-dynamical D$p$-branes.
For this, we applied the boundary state formalism and
the general form of the vertex operator,
corresponding to a massless closed string.
Due to the presence of the various parameters
in the setup, the value of the amplitude can be
accurately modified to any desirable value.

We changed the foregoing amplitude
for the branes with a large distance.
For this case, we observed that one of the
following radiations can potentially occur:
two radiations from the branes and another
one between the branes. However,
all of these three possible processes can
happen for the graviton and dilaton emissions,
while the Kalb-Ramond radiation cannot happen
between the branes.


\end{document}